\documentclass[preprint,pre,floats,aps,eqsecnum,tightenlines,amsthm,amsmath,amssymb,amsfonts]{revtex4} 

\usepackage{inputenc,fontenc,lineno}
\usepackage{xcolor,color,bm}
\usepackage{ulem}
\usepackage{graphicx}
\usepackage{appendix}
\usepackage{mathtools}

\newcommand{\mean}[1]{\left\langle#1\right\rangle}

\begin{document}

\title{Spectrum correction in Ekman-Navier-Stokes turbulence}

\author{V.J. Valad\~{a}o$^{1}${\footnote{Corresponding author: victor.dejesusvaladao@unito.it}}, G. Boffetta$^{1}$, F. De Lillo$^{1}$, S. Musacchio$^{1}$, and M. Crialesi-Esposito$^{2}$}
\affiliation{$^{1}$Dipartimento di Fisica and INFN - Università degli Studi di Torino, Via Pietro Giuria, 1, 10125 Torino TO, Italy.}
\affiliation{$^{2}$DIEF, University of Modena and Reggio Emilia, 41125 Modena, Italy.}
\vspace{1.5cm}

\begin{abstract}
  The presence of a linear friction drag affects significantly
  the dynamics of turbulent flows in two-dimensions.
  At small scales, it induces a correction to the slope of the energy spectrum
  in the range of wavenumbers corresponding to the direct enstrophy cascade.    
  Simple arguments predict that this correction is proportional to the ratio of the friction coefficient
  to the characteristic deformation rate of the flow. 
  In this work, we examine this phenomenon by means of a set of GPU-accelerated numerical simulations at high
  resolutions, varying both the Reynolds number and the friction coefficient.
  Exploiting the relation between the energy spectrum and the enstrophy flux,
  we obtain accurate measurements of the spectral scaling exponents.  
  Our results show that the exponent of the spectral correction follows a universal linear law
  in which the friction coefficient is rescaled by the enstrophy injection rate.
  
  \vspace{0.5cm}
  \noindent \textbf{Keywords:} 2D Turbulence, Direct Enstrophy Cascade, Ekman friction
\end{abstract}

{\let\clearpage\relax\maketitle\newpage}

\section{Introduction}
\label{introduction}

A significant number of natural fluid dynamical systems, such as atmospheric
jet streams~\cite{vallis2017atmospheric}, ocean 
currents~\cite{lapeyre2006dynamics}, and planetary
flows~\cite{siegelman2022moist}, exhibit turbulent motion. 
While the turbulent flow is typically a three-dimensional
(3D) phenomenon, many real-world phenomena display characteristics of
two-dimensional (2D) turbulence across various scales. Examples of such 2D
turbulence include large-scale patterns in the Earth's
atmosphere~\cite{juckes1994quasigeostrophic}, flows confined within thin fluid layers by geometric
boundaries~\cite{xia2009spectrally,benavides2017critical,musacchio2017split,musacchio2019condensate,zhu2023circulation}
and the behaviour of conducting fluids under strong magnetic
fields~\cite{moffatt1978magnetic}. 
In contrast to 3D turbulence, where energy is transferred from larger to
smaller scales in a forward
cascade~\cite{kolmogorov1941local,frisch1995turbulence}, 2D turbulence is
characterised by a dual cascade: an inverse energy cascade towards large
scales and a direct cascade in which the enstrophy is transferred towards small
scales~\cite{kraichnan1971inertial,boffetta2012two}.

Linear friction, also referred to as Ekman friction, is commonly 
added to the Navier-Stokes (NS) equations in 2D
as an essential ingredient for modelling real-world effects such as
boundary layer dynamics, bottom drag in oceans, atmospheric resistance~\cite{vallis2017atmospheric}
or the air friction in experiments with of soap films~\cite{rivera2000external}.
Linear friction has an important role in the process of the inverse energy cascade since it 
provides a sink for the energy transferred to large scale, allowing to attain a statistically stationary state \cite{kraichnan1967inertial,leith1968diffusion,batchelor1969computation,chertkov2007dynamics,boffetta2012two}.
The presence of linear damping significantly affects also 
the statistical properties of the direct enstrophy cascade.
Theoretical investigations \cite{nam2000lagrangian,bernard2000influence}
and numerical simulations \cite{boffetta2002intermittency}
of the Ekman-Navier-Stokes equations have shown that the dissipation
of enstrophy due to the friction at small scales causes a steepening
of the energy spectrum. This results in a correction $\xi >0$ to the scaling exponent of the
spectrum $E(k) \sim k^{-(3+\xi)}$ with respect to the Kraichnan prediction
for the direct enstrophy cascade~\cite{kraichnan1971inertial}.
Theoretical arguments based on the similarities between the process of the direct enstrophy cascade
and the chaotic advection of passive scalar fields \cite{nam2000lagrangian,bernard2000influence,boffetta2002intermittency}
have shown that the correction $\xi$ is determined by the statistics of the stretching rates of the flow
and it is proportional to the friction coefficient.
More generally, these studies have shown that the friction drag causes the breakdown of self-similar
scaling of the vorticity structure functions in the range of scales of the direct enstrophy cascade,
resulting in anomalous scaling exponents which depend on the friction coefficient~\cite{nam2000lagrangian,bernard2000influence}.
These relationships provide an intriguing link between the chaoticity of Lagrangian trajectories
and the statistical scaling laws in 2D turbulent flows. 

In this paper, we pursue the investigation of the effects of linear friction on the direct enstrophy cascade
in 2D turbulence by means of a set of high-resolution numerical simulations
of the NS equations varying the Reynolds number and friction coefficient.
These achievements are made possible by the development of a numerical code specifically designed
for single Graphics Processing Unit (GPU) which allows us to greatly 
speed up the simulation with respect to traditional CPU-based methods. 
Our findings provide deeper insights into the relationship between the friction and the slope of the energy spectrum
showing that the correction $\xi$ displays a universal linear dependence
as a function of the friction coefficient rescaled by the characteristic time-scale based on the enstrophy injection rate.
We tested the robustness of these results by allowing the development of the inverse energy cascade
for the simulations with the largest Reynolds number.
We also show that fitting the power-law behaviour of the enstrophy flux instead of the energy spectrum provides a 
more accurate measurement of the correction $\xi$.
This method overcomes the difficulties arising from the presence of a logarithmic correction to the spectrum
which affects the direct measurement of the correction $\xi$ in the limit of vanishing friction. 

The paper is organised as follows: Sec.~II provides an overview of the
phenomenology of the direct enstrophy cascade in the presence of a linear friction drag,
both in the limit of vanishing friction and with 
finite friction. The results of numerical simulations 
are discussed in Sec.~III. Finally, Sec.~IV discusses the implications of our
findings and suggests directions for future research. Details on the
pseudospectral method and performance benchmarks of our simulations are shown
on Appendix~\ref{appa}. In Appendix~\ref{appb}, we discuss the difficulties of retrieving the scaling exponent
from direct measurement of the slope of the energy spectrum, especially in the frictionless limit. 

\section{Direct enstrophy cascade in 2D turbulence}
\label{sec2}

The dynamics of an incompressible velocity field $\bm{u}(\bm{x},t)$
in two dimensions can be conveniently written in terms of the vorticity field 
$\omega(\bm{x},t)=\partial_x u_y - \partial_y u_x$ as
\begin{equation}\label{eq2.2}
\partial_t\omega+\bm{u}\cdot\bm{\nabla}\omega=\nu\nabla^2\omega-\mu\omega+f\ ,\
\end{equation}
where $\nu$ is the kinematic viscosity (with units of length squared over time) and $\mu$ is the friction coefficient (an inverse time).
The forcing term $f(\bm{x},t)= \partial_x F_y - \partial_y F_x $ 
(inverse time squared) is related to the  
external force field $\bm{F}(\bm{x},t)$ which sustains the flow.
The forcing field is assumed to
be random with a characteristic spatial correlation length of $\ell_f$.

In the inviscid, frictionless, unforced limit, the model (\ref{eq2.2}) conserves the kinetic
energy $E=\mean{|{\bm u}|^2}/2$ and the enstrophy $Z=\mean{\omega^2}/2$, 
where the brackets $\mean{(.)}$ indicate the spatial average. 
In the presence of forcing and dissipation, the energy and enstrophy balances read:
\begin{equation}
\frac{dE}{dt} = -2 \nu Z - 2 \mu E + \mean{{\bm u} \cdot {\bm F}} =
-\varepsilon_{\nu}-\varepsilon_{\mu}+\varepsilon_I \ ,\
\label{2.4}
\end{equation}
and
\begin{equation}
\frac{dZ}{dt} = -2 \nu P - 2 \mu Z + \mean{\omega f} =
-\eta_{\nu}-\eta_{\mu}+\eta_I \ ,\
\label{eq2.5}
\end{equation}
where $P=\mean{|\bm\nabla\omega|^2}/2$ is the so-called palinstrophy
that controls the viscous dissipation of enstrophy.

The different terms in (\ref{2.4}-\ref{eq2.5}) define,
together with the characteristic scales of the forcing 
$\ell_f=2\pi\sqrt{\varepsilon_I/\eta_I}$, the viscous dissipation scale
$\ell_{\nu}=2\pi\sqrt{\varepsilon_{\nu}/\eta_{\nu}}$ and the friction scale
$\ell_{\mu}=2\pi\sqrt{\varepsilon_{\mu}/\eta_{\mu}}$. When these scales are 
well separated, $\ell_{\nu} \ll \ell_f \ll \ell_{\mu}$, one expects the 
development of a direct enstrophy cascade in the inertial range of scales 
$\ell_{\nu} \ll \ell \ll \ell_f$ and an inverse energy cascade in the scales 
$\ell_f \ll \ell \ll \ell_{\mu}$ \cite{boffetta2012two}.

The central statistical object in the classical theory of turbulence is the
energy spectrum $E(k)$ defined as $\int E(k) dk = E$ or, equivalently,
as $\int k^2 E(k) dk = Z$.
The spectral flux of enstrophy $\Pi_{Z}(k)$ in the direct enstrophy cascade 
can be related to the energy spectrum according
to the following dimensional closure \cite{boffetta2012two}
\begin{equation}
\Pi_{Z}(k) = \lambda_k E(k) k^3 \ .\
\label{eq2.6}
\end{equation}
In (\ref{eq2.6}), $\lambda_k$ represents the characteristic frequency of 
deformation of the eddies at wavenumber $k$ which can be 
expressed in terms of the energy spectrum as
\begin{equation}
\lambda_k^2 = \int_{k_{min}}^{k} E(p) p^2 dp \ ,\
\label{eq2.7}
\end{equation}
where $k_{min} = 2\pi/\ell_\mu$ is the minimum wavenumber associated 
with the largest scale of the flow $\ell_\mu$. The upper limit in the 
integral reflects the fact that the fluid motion at scales smaller 
than $1/k$ acts incoherently, and therefore its average contribution 
to the deformation rate of the eddies of size $1/k$ vanishes. 
Considering a scale-invariant energy spectrum $E(k) \propto k^{-\beta}$,
the integral in (\ref{eq2.7}) is dominated by the upper limit $k$ provided
that the scaling exponent is in the range $\beta < 3$, satisfying the locality condition
\cite{frisch1995turbulence}, and therefore $\lambda_k \simeq E(k)^{1/2} k^{3/2}$.

In the absence of friction ($\mu = 0$), one can assume that the forcing
and dissipation terms are both negligible in the enstrophy inertial 
range, and therefore the flux of enstrophy is constant, $\Pi_{Z}(k) = \eta$.
This assumption, in combination with the dimensional relation (\ref{eq2.6}) gives the prediction
$E(k) \simeq \eta^{2/3} k^{-3}$. However, this result is not self-consistent
because the spectral exponent $\beta=3$ is at the border of locality. 
According to Equation~(\ref{eq2.7}), this gives a logarithmic correction for the $\lambda_k$
and consequently, a non-constant, log-dependent enstrophy flux. 
A solution to this problem was already proposed by Kraichnan
\cite{kraichnan1971inertial}. By taking the derivative of (\ref{eq2.7}) 
and plugging in (\ref{eq2.6}) one obtains
\begin{equation}
\Pi_Z(k) = 2 k \lambda_k^2 \frac{d \lambda_k}{dk} 
\label{eq2.7bis}
\end{equation}
from which, assuming a constant enstrophy flux 
$\Pi_{Z}(k)=\eta$, one obtains a log-dependent deformation frequency
\begin{equation}
\lambda_k = \left( \frac{3}{2} \eta \ln \left(\frac{k}{k_f}\right) \right)^{1/3} \ .\
\label{eq2.7tris}
\end{equation}
Using this expression in (\ref{eq2.6}) one ends with the prediction
\cite{kraichnan1971inertial}
\begin{equation}
E(k) \simeq \eta^{2/3} k^{-3} \left[\ln(k/k_{f})\right]^{-1/3} \ .\
\label{eq2.8}
\end{equation}

The presence of friction drag changes significantly the whole process of 
the enstrophy cascade. In particular, it excludes the possibility of a 
constant flux of enstrophy, causing a steepening of the energy spectrum 
\cite{nam2000lagrangian,bernard2000influence}.
This phenomenon can be explained by a simple argument.
In the presence of linear friction, from \eqref{eq2.2}, one has the 
following expression for the rate of enstrophy transfer \cite{boffetta2012two}
\begin{equation}
\frac{d\Pi_Z(k)}{dk} = - 2\mu k^2 E(k)
\label{eq2.9}
\end{equation}
which states that part of the flux is removed in the cascade at a 
rate proportional to the friction coefficient $\mu$.
This causes the steepening of the energy spectrum, with a spectral slope 
$\beta > 3$ which exceeds the range of locality. 
As a consequence, the integral (\ref{eq2.7}) 
is dominated by the contribution of the wavenumbers $k_{min} \le k \le k_f$,
while the contribution of the wavenumbers $k > k_f$ is negligible,
resulting in a constant deformation rate $\lambda_{k}=\lambda$. 
Using this assumption in Equation~(\ref{eq2.6}), one immediately obtains the 
solution
\begin{equation}
E(k) \simeq \frac{\eta}{\lambda} k^{-3} (k/k_f)^{-\xi}
\label{eq2.10}
\end{equation}
with the correction to the dimensional scaling exponent
\begin{equation}
\xi = \frac{2\mu}{\lambda} \, .
\label{eq2.11}
\end{equation}

We remark that the above argument can be made more rigorous in the 
physical space where the role of $\lambda$ is replaced by the stretching rate
of the smooth, chaotic flow. By taking into account the finite-time 
fluctuations of the stretching rates, one predicts the breakdown of
self-similar scaling and 
the production of intermittency in the statistics of the vorticity 
field \cite{nam2000lagrangian} which has been observed in numerical 
simulations \cite{boffetta2002intermittency}.

\section{Numerical simulations of the direct cascade with friction}
\label{sec3}

We tested the prediction of the previous section, in particular the
correction (\ref{eq2.10}) to the energy spectrum in the presence of friction 
by means of extensive direct numerical simulations of Equation~\eqref{eq2.2} at very
high resolutions (up to $16384^2$ gridpoints). To accomplish this, we used a pseudo-spectral code implemented 
on a single GPU. A detailed discussion of the code and its performance can be 
found in Appendix~\ref{appa}.
\begin{figure}[ht]
\centering
\includegraphics[width=.65\textwidth]{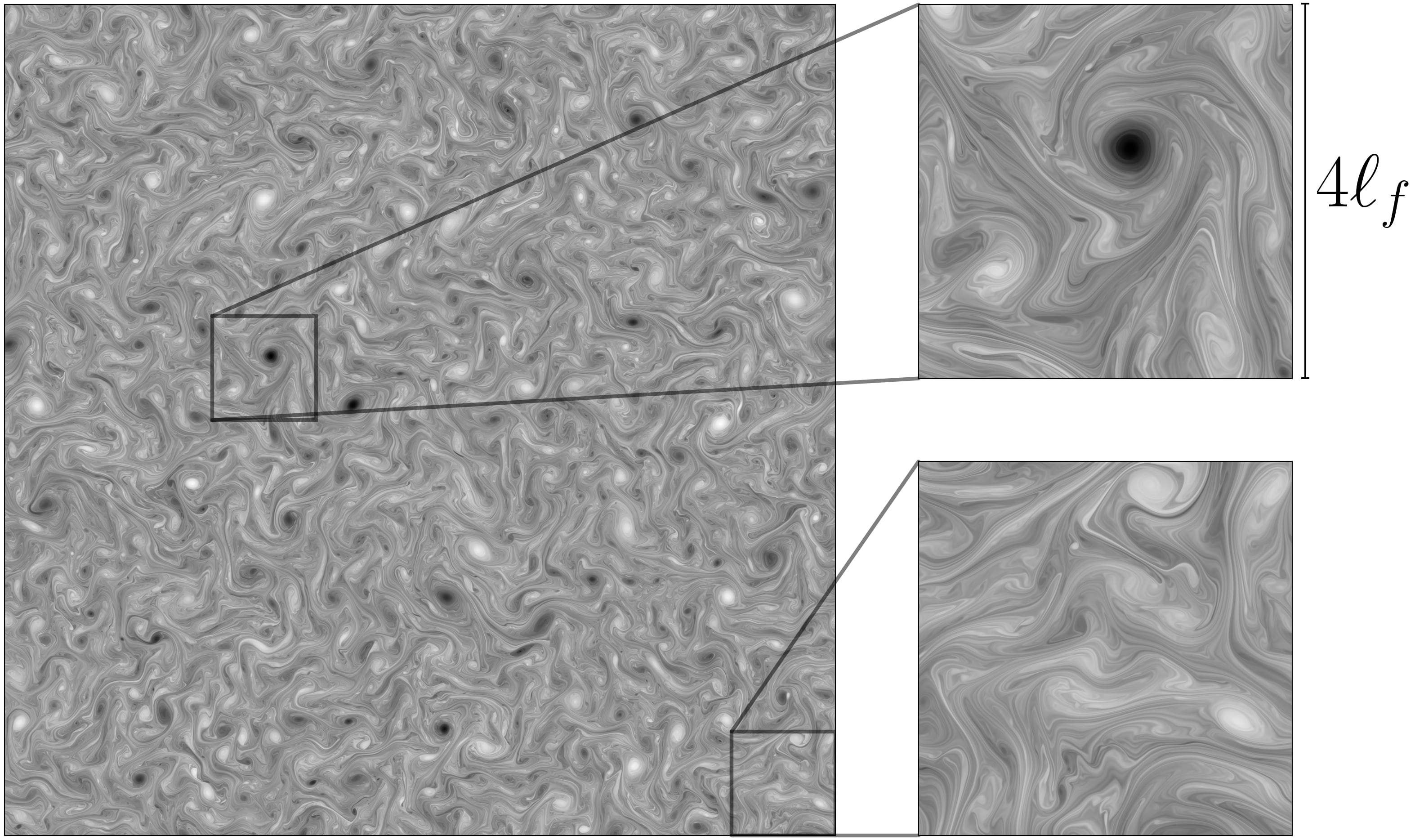}
\caption{Snapshot of $\omega(\bm x)$ for Run C with
$\mu\eta_I^{-1/3}\approx0.01$. The upper panel illustrates a region where the
flow is dominated by a single, large vortex, approximately the size of the
forcing scale. The lower panel depicts a region with no dominant vortices.
}
\label{fig1}
\end{figure}

Simulations are done in a square box of size $L_x=L_y=2\pi$, with regular grid
of resolution $N=N_x=N_y$.
The turbulent flow is sustained by a Gaussian random forcing $f(\bm x,t)$
with zero mean and white-in-time correlations, acting in a
narrow spherical shell of thickness $\Delta k$ 
centred at $k_f$ in the wavenumber space.
Such a forcing provides an average energy and enstrophy injection rate, 
$\varepsilon_I$ and $\eta_I$, respectively, that are related by 
$\eta_I \approx \varepsilon_I k_f^2$ when $\Delta k \ll k_f$.

Three sets of simulations have been done with different
resolutions and viscosities $\nu$, each one covering a large range of friction
coefficients $\mu$. 
\begin{table}[!t]
\begin{center}
\begin{tabular}{ ||c||c|c|c|c|c|c|c||}
 \hline
 \hline
 Run & $N$ & $\nu$ & $k_f\pm \Delta k$ & $\eta_I$ & $Re_\nu $ & $k_{\max}\ell_\nu$ & $\mu\times10^{2}$\\
 \hline
 \hline
 A & 4096  & $2\times10^{-5}$    & $8\pm1$  &   9.615 & 65584  & 4.19 & 1,4,7,10,20,30,40,50,60,80      \\
 B & 8192  & $5\times10^{-6}$    & $16\pm1$ &  34.560 & 100463 & 3.38 & 4,6,10,20,30,40,50,60,80,100 \\
 C & 16384 & $1.25\times10^{-6}$ & $32\pm1$ & 114.750 & 149877 & 2.77 & 6,12,18,36,48,60,72,96,120               \\
 \hline
 \hline
\end{tabular} 
\end{center}
\caption{The most relevant parameters of the simulation include $k_f = 2\pi/\ell_f$ and $k_{\max} = N/3$, since we use the 2/3 de-aliasing method. The viscous scale and the Reynolds number are given by $\ell_\nu = \nu^{1/2} \eta_I^{-1/6}$ and $Re = \left(\ell_\nu / \ell_f\right)^2$. Both should be taken as lower bound estimates, since if one considers $\eta_\nu$ instead of $\eta_I$ as the proper dimensional quantity, one obtains strictly larger values for $\ell_\nu$ and $Re$, since $\eta_\nu < \eta_I$.}
\label{table1}
\end{table}
By increasing the resolution, we increased the forcing scale to allow 
the development of a narrow inverse cascade in the simulations at the 
highest resolution at low friction.
Table \ref{table1} reports the most relevant parameters of our simulations in arbitrary units.
In all cases, small scales are well resolved ($k_{\max} \ell_\nu \ge 2.77$). 
We remark that since the forcing amplitude is kept constant, the 
enstrophy injection rate increases with the forcing wavenumber and therefore
with the resolution. 

Figure~\ref{fig1} shows a snapshot of the vorticity field taken from 
Run $C$, the highest resolution. The size of the vortices observed in the flow
corresponds to the forcing scale, as shown in the upper right panel, which is
reduced by increasing the resolution, as indicated in Table~\ref{table1}.
\begin{figure}[!th]
\centering
\includegraphics[width=.6\textwidth]{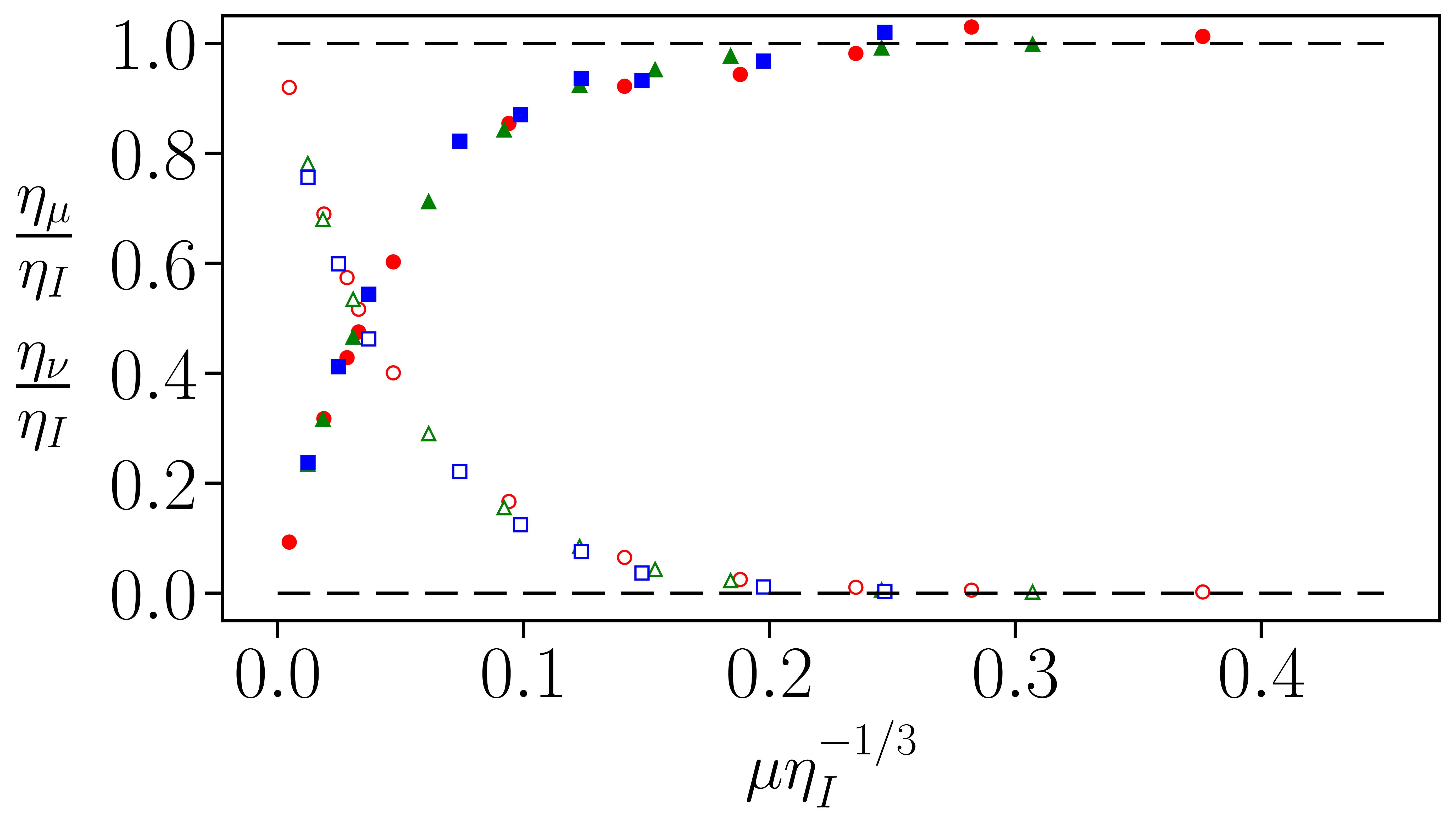}
\caption{The ratio of friction dissipation $\eta_{\mu}$ (filled symbols) and viscous dissipation $\eta_{\nu}$ (open symbols) to the enstrophy input $\eta_I$ for the Runs A, B, and C are represented by red circles, green triangles, and blue squares, respectively. The friction coefficient is made nondimensional using the timescale associated with $\eta_I$.}
\label{fig2}
\end{figure}

The enstrophy balance (\ref{eq2.5}) is shown in Figure~\ref{fig2} for all the
simulations in stationary conditions.
Remarkably, the curves at different inputs and dissipations collapse when the 
friction coefficient is made 
dimensionless with the time-scale associated with the enstrophy injection,
i.e. $\tau_I=\eta_I^{-1/3}$. 
Moreover, we observe from Figure~\ref{fig2} that for $\mu \eta_I^{-1/3} \gtrsim 0.2$,
the viscous dissipation is negligible, and the entire enstrophy flux that cascades 
towards small scales is dissipated by friction before reaching the viscous scale.
\begin{figure}[!th]
\centering
\includegraphics[width=.98\textwidth]{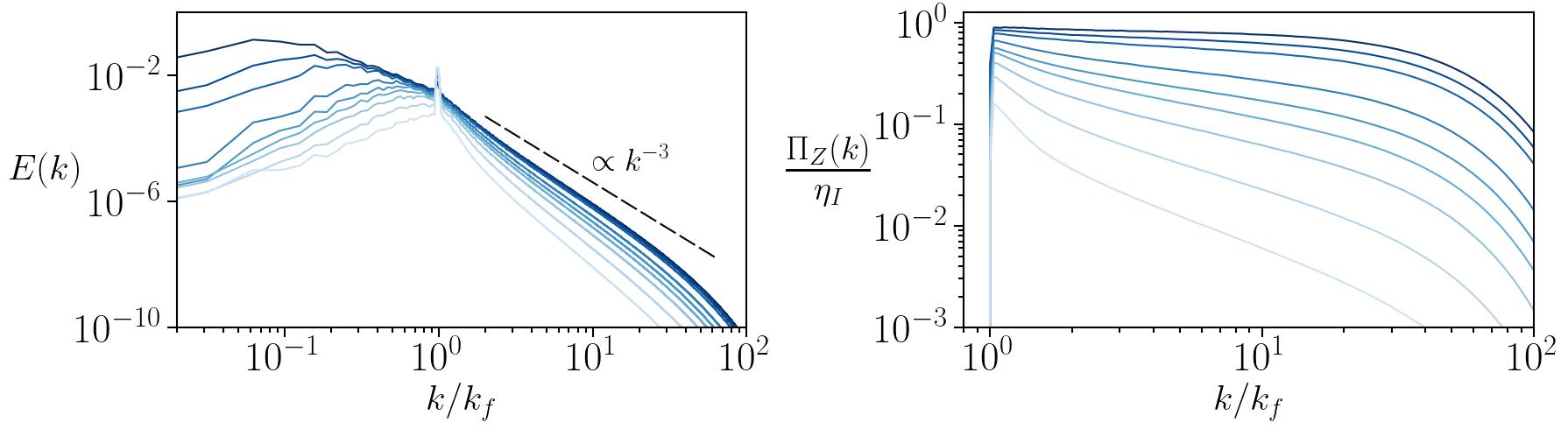}
\caption{The left panel shows the energy spectra in arbitrary units, while the right panel shows the enstrophy fluxes, both for the simulations of Run C. Darker curves correspond to smaller values of $\mu$.}
\label{fig3}
\end{figure}

Figure~\ref{fig3} (left) shows the time-averaged energy spectra for the different
simulations of Run C. In all cases, in the direct cascade range, the spectrum
shows a power-law scaling steeper than the simple dimensional prediction
$E(k)\propto k^{-3}$ with increasing scaling exponent $\beta$ for 
larger friction $\mu$, as expected. 
The three darker curves, corresponding to smaller friction values, 
display a short inverse cascade at wavenumber $k<k_f$ with an exponent 
close to the dimensional prediction for the energy cascade, $k^{-5/3}$. 

From Figure~\ref{fig3}, it is clear that fitting the scaling exponent $\beta$ directly 
from the spectrum is problematic due to the presence of the peak corresponding to the 
forcing wavenumber.
Moreover, in the limit $\mu \to 0$, the energy spectrum has the logarithmic 
correction \eqref{eq2.8} to the power-law scaling, and we can expect this to 
persist for small values of the friction coefficient $\mu$.
Indeed, we found empirically that simply fitting the spectra with a
power law exponent $3+\xi$ does not correctly recover
the limit $\xi=0$ for vanishing friction (see Appendix~\ref{appb}).
\begin{figure}[!th]
\centering
\includegraphics[width=.6\textwidth]{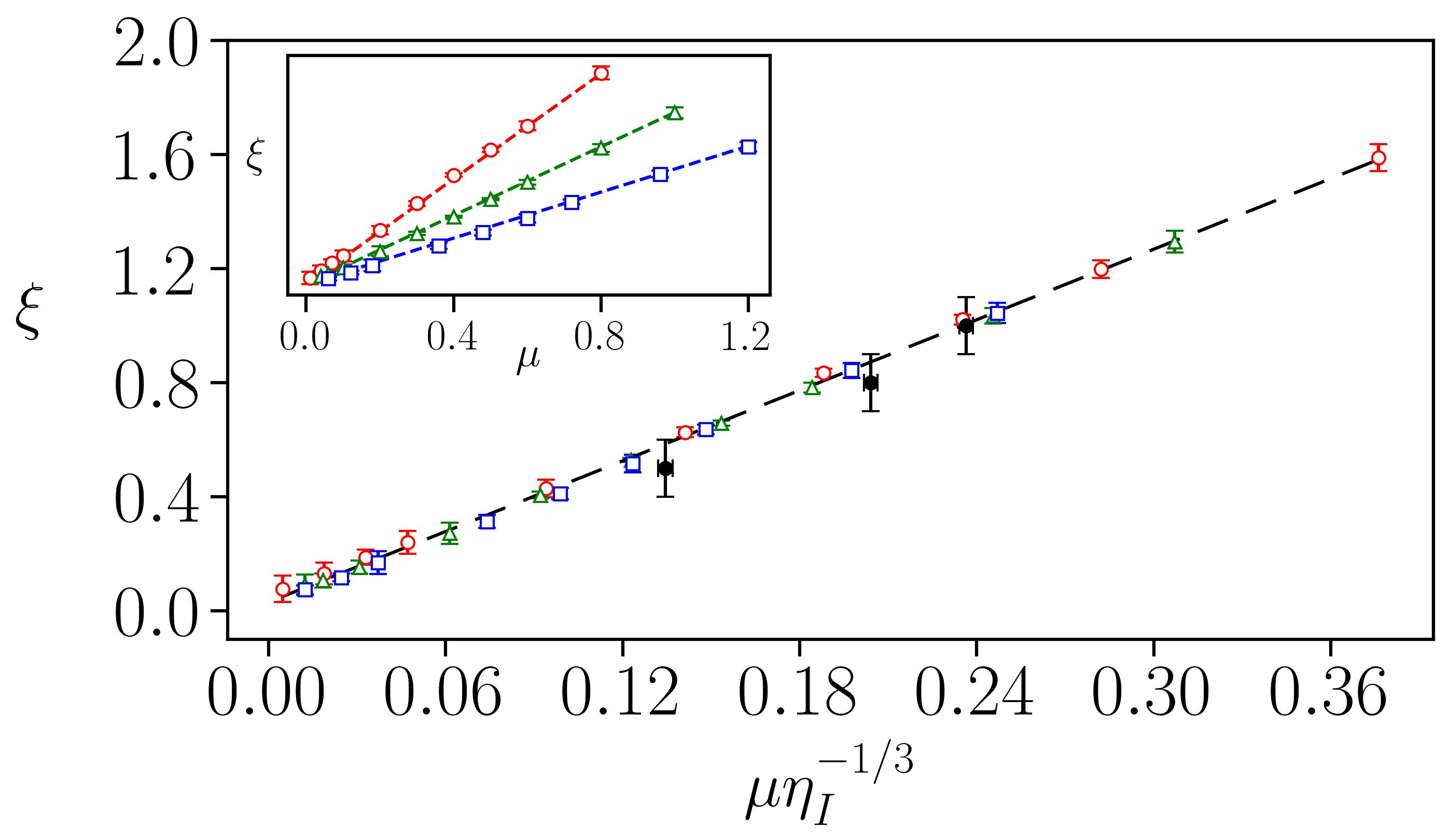}
\caption{Spectral correction $\xi$ as a function $\mu\eta_I^{-1/3}$
for Runs A (open circles), B (open triangles), and C (open squares). 
Filled symbols represent experimental data from \cite{boffetta2005effects}. 
The dashed line represents the relation $\xi=a\big(\mu\eta_I^{-1/3}\big)+b$
where $a=4.1\pm0.3$ and $b=0.03\pm0.05$. Inset: spectral correction
$\xi$ as a function of the dimensional $\mu$ in arbitrary units for all simulations.
Error bars are estimated by varying the fitting range $k_0\in[3k_f,5k_f]$ and
$k_1\in[7k_f,9k_f]$.}
\label{fig4}
\end{figure}

To overcome these difficulties, we decided to measure the correction 
$\xi(\mu)$ directly from the power-law scaling of the flux $\Pi(k)$, 
since the two quantities are related by (\ref{eq2.9}). From the theory,
we do not expect logarithmic corrections in the enstrophy flux. 
The right panel of Figure~\ref{fig3} shows the spectral enstrophy fluxes 
for Run C, and we observe a clear power-law scaling in an 
intermediate range of wavenumbers $k \in [k_0, k_1]$ (with $k_0 \simeq 3 k_f$
and $k_1 \simeq 9 k_f$), far from the forcing and dissipation scales.

The spectral correction $\xi$ obtained from the power-law fit of the 
spectral fluxes in the range $k \in [k_0, k_1]$ is shown in Figure~\ref{fig4}.
From the inset of Figure~\ref{fig4}, it is evident that the correction $\xi(\mu)$
is proportional to the friction coefficient $\mu$, as predicted by 
(\ref{eq2.11}), with a different slope for the different Runs 
characterised by varying input. We find that, once again, the correct 
timescale for non-dimensionalizing the friction parameter is based on the
enstrophy input rate. Indeed, as shown in Figure~\ref{fig4}, when plotted 
as a function of the dimensionless friction coefficient $\mu \eta_I^{-1/3}$,
all the data from the different Runs collapse onto a single line. Moreover, 
in the limit $\mu \to 0$, the spectral correction fitted by the collapsed 
curve is compatible with zero.

We also show in Figure~\ref{fig4}, the results of an experiment 
in a thin layer of conducting fluid where the spectral correction
has been measured \cite{boffetta2005effects}.
The experimental setup is a square tank of side $L=50$ cm 
filled with a fluid of thickness $h=0.8$$-$$1.0$ cm that provides different
bottom friction coefficient varying in the range of 
$\mu=0.037$$-$$0.069$ s$^{-1}$.
The typical velocities and vorticities measured are in the range
$v_{rms}=0.79$$-$$1.33$ cm/s and $\omega_{rms}=0.60$$-$$0.75$ s$^{-1}$,
respectively. The agreement between experimental data and simulations again
supports that the spectral correction $\xi$ depends on the rescaled
variable $\mu\eta_I^{-1/3}$ only.
We remark that the above rescaling is not 
the only possibility: one could use $Z^{1/2}$ as an inverse time of the 
flow, but this would not lead to the data collapsing as shown in Figure 4.

\section{Conclusions}
\label{conclusion}

In this study, we examined the effects of a linear friction on the direct
enstrophy cascade in 2D turbulence using high-resolution numerical simulations
of the Navier-Stokes equations. 
By profiting from a GPU-accelerated code, we explored a
wide range of Reynolds numbers and friction coefficients $\mu$ uncovering key
insights into the dynamics of 2D turbulent flows with linear damping.

Our results confirm that the linear friction introduces a correction 
$\xi$ to the scaling exponent of the energy spectrum in the direct enstrophy 
cascade, steepening the dimensionally predicted slope. 
Theoretically, this correction is expected to scale proportionally 
to the ratio $\mu/\lambda$ where $\lambda$ is the average deformation
rate of the flow.
Our simulations confirmed the scaling $\xi\propto\mu$,
providing robust evidence supporting this scaling across a broad parameter
space, including different forcing scales, friction coefficients
and Reynolds numbers. 
A precise measure of the correction $\xi$ is obtained from the scaling 
law of the enstrophy flux and exploiting its relation with the energy 
spectrum which, in turn, gives less precise results in particular for small 
values of the friction coefficient.
By this procedure, we find that a consistent measure of the deformation rate, 
in the range of parameters explored here, is expressed in terms of the 
enstrophy input rate $\eta_I$ and that $\lambda \propto \eta_I^{1/3}$.
This latter result is strongly supported by the comparison of our data to
the experimental result from \cite{boffetta2005effects}.

The present result is obtained in a regime of moderate friction,
given by the dimensionless coefficient $\mu \eta_I^{-1/3}<1$. 
It would be interesting to extend this study to the opposite regime 
$\mu \eta_I^{-1/3}>1$, where friction directly affects 
the statistics of the velocity field at the forcing scale. 
In such a regime, it is expected that the deformation rate $\lambda$
depends on $\mu$ and therefore we expect a non-linear scaling of 
the spectral coefficient $\xi$ on the parameters. 

\section*{Acknowledgement}

This work has been supported by Italian Research Center on High Performance Computing Big Data and Quantum Computing (ICSC), project funded by European Union - NextGenerationEU - and National Recovery and Resilience Plan (NRRP) - Mission 4 Component 2 within the activities of Spoke 3 (Astrophysics and Cosmos Observations). We acknowledge HPC CINECA for computing resources within the INFN-CINECA Grant INFN24-fldturb.

\appendix

\section{Numerical integration of generalised 2D turbulence models}
\label{appa}

To build numerical solvers for a broader class of turbulent models, we rewrite \eqref{eq2.2} in a more general formulation,
\begin{equation}
(\partial_t+\mathcal{L}_{\nu,\mu}^{n,m})\omega+J(\omega,\psi)=f\ ,\
\label{eqa.1}
\end{equation}
where we introduced a generalised linear dissipative operator,
\begin{equation}
\label{eqa.2}
\mathcal{L}_{\nu,\mu}^{n,m}\equiv(-1)^n\nu_{2n} \nabla^{2(n+1)}+
(-1)^m\mu_{2m} \nabla^{-2m} \ ,\
\end{equation} 
representing a positive-diagonal operator in the Fourier space $\hat{\mathcal{L}}_{\nu,\mu}^{n,m}(k)=\nu_{2n}k^{2(n+1)}+\mu_{2m} k^{-2m}$. Although this paper is devoted to the study of the direct cascade in 2D NS turbulence, the equation (\ref{eqa.1}) contains a whole class of turbulence models known as $\alpha$-turbulence \cite{pierrehumbert1994spectra}. The definition of this class of model is better understood through the relation between the generalised vorticity $\omega({\bm x},t)$ and the stream function $\psi({\bm x},t)$, represented in the Fourier space through
\begin{equation}
\label{eqa.3}
\hat{\omega}({\bm k},t)=|{\bm k}|^{\alpha}\hat{\psi}({\bm k},t)\ .\
\end{equation}
In the following, we will discuss the case $\alpha=2$ but the scheme can be adapted to any value of $\alpha$.

The generalised dissipative operator has the role discussed in Section~\ref{sec2}, i.e. to provide stationary states and prevent condensate formations. For $m=n=0$ one recovers the standard friction/viscosity terms. In comparison, for $m,n>0$ depending on the orders $n$ and $m$ of the dissipative operator, the coefficients $\mu$ and $\nu$ have different dimensional roles and can dissipate over a more narrow range of scales. For example, hyperviscosity ($n>0$) is used to diminish the action of dissipation on the dissipative subrange, leading to extended inertial ranges at the cost of a bigger thermalisation effect (bottleneck) of high wavenumber \cite{haugen2004inertial,frisch2008hyperviscosity}. Moreover, one reason to introduce hypofriction ($m>0$) instead of normal friction is to avoid the correction to the enstrophy cascade discussed in Section~\ref{sec2}.

We developed and tested an original pseudospectral code to integrate the general model on Nvidia hardware. Pseudospectral schemes are widely used in numerical studies of turbulence because of their accuracy in derivatives and the simplicity of inverting the Laplace equation. Another practical advantage is that most of the resources in the pseudospectral scheme are used to compute the Fast Fourier Transforms (FFT) necessary to move back and forth from Fourier space (where derivatives are computed) to physical space (where products and other nonlinear terms are evaluated). Therefore, to make the code efficient for a given architecture, it is (almost) sufficient to have an efficient FFT. 

The numerical code gTurbo2D uses a standard Runge-Kutta (RK) scheme to time advance the solution with exact integration of the linear terms. In the simple case of a second-order RK scheme, the evolution of the vorticity field in (\ref{eqa.1}) from the time $t$ to $t+dt$, with $dt$ the timestep of the simulation is given by 
\begin{equation}
\hat{\omega}({\bm k},t+dt) =
e^{-\hat{\mathcal{L}}dt} \hat{\omega}({\bm k},t)+
e^{-\hat{\mathcal{L}}dt/2}
\hat{N}\left(e^{-\hat{\mathcal{L}}dt/2} \hat{\omega}'\right) dt
\label{eqa.4}
\end{equation}
where $\hat{(.)}$ represents the Fourier transformed fields and
\begin{equation}
\hat{\omega}'=\hat{\omega}({\bm k},t)+
\hat{N}\left(\hat{\omega}({\bm k},t)\right) dt/2 \ .\
\label{eqa.5}
\end{equation}
It is worth emphasising that the timestep $dt\ll dx/U_{rms}$, where $dx=2\pi/N$ and $U_{rms}=\sqrt{\mean{|\bm v|^2}}$ is the root-mean-square of the velocity field generated by $\omega$. Such a CFL condition depends strongly on the stability of the time integration scheme. In our simulations on the main text, $U_{rms} dt/dx$ is always smaller than $1/10$.

The evaluation of the nonlinear term $\hat{N}$ is partially done in the physical space (to avoid the computation of convolutions). In the present implementation of the code, the evaluation of the nonlinear term is done as follows. From the vorticity field in Fourier space, the code computes the stream function by inverting (\ref{eqa.3}). The two components of the velocity $\hat{v}_i$ are then obtained from the derivatives of $\hat{\psi}$ and then transformed in the physical space together with the vorticity (this step requires 3 inverse FFTs). The products $(v_i \omega)$ are computed (and stored in the same arrays of the velocity) and transformed back in Fourier space (this requires 2 direct FFTs). Finally, the divergence of $\hat{(v_i \omega)}$ is computed and stored in the original array. Therefore the evaluation of the nonlinear term requires 5 FFTs and each step of the n-order RK scheme requires $5n$ FFTs. 

The code gTurbo2D is written in Fortran 90 with OpenACC, which enables the use of Nvidia hardware through compiler directives. For the FFTs, the code makes massive use of the CUDA FFT library, compatible with the OpenACC programming paradigm. Simulations are performed on {\it Leonardo} machine, a pre-exascale Tier-0 supercomputer where, each of the 3456 computing nodes is composed of a single-socket processor of 32-core at 2.60GHz, 512 GB of RAM and, 4 Nvidia A100 GPUs of 64GB each connected by NVLink 3.0. The version of gTurbo2D used for this work is a single GPU code while the multi-GPU version is under development. We remark that the study of 2D turbulence requires much less memory than 3D (a single scalar field in two dimensions) and the remarkable resolution of $N^2=32768^2$ grid points can be reached on a single GPU. However, large resolutions require very small time steps and therefore the resolution is limited not only by the memory but also by the speed of the code. 
\begin{figure}[ht]
    \includegraphics[width=.98\textwidth]{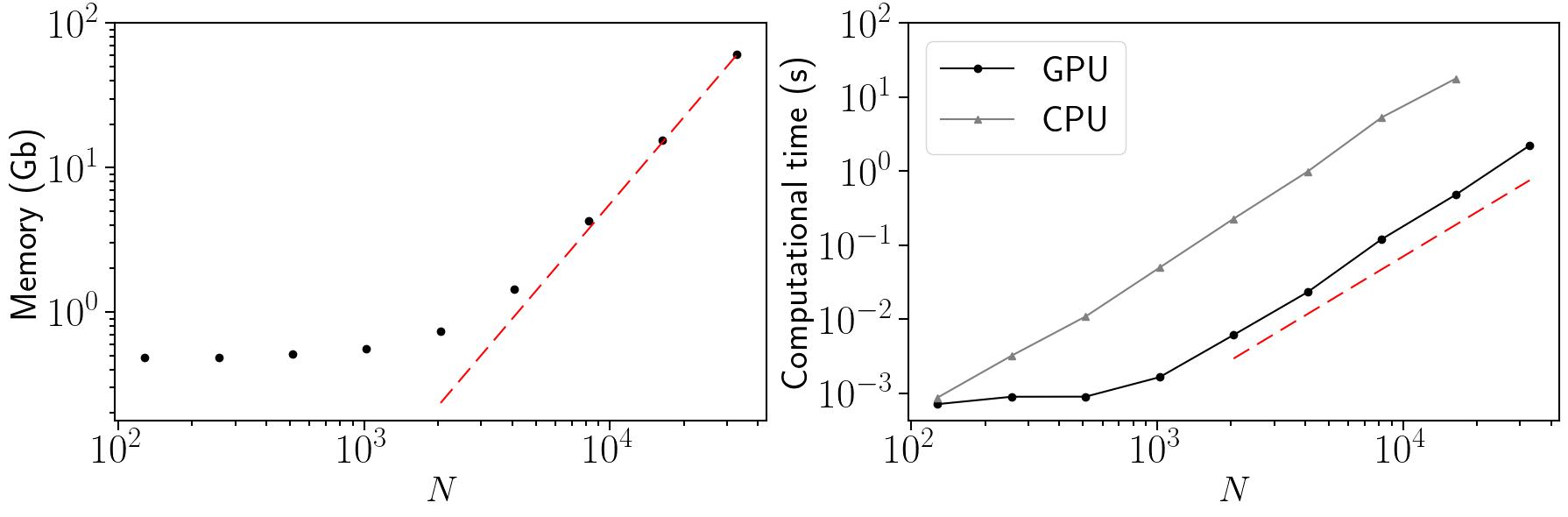}
    \caption{Left panel shows GPU's memory usage while right panel shows mean elapsed time (computed with $1000$ timesteps) as functions of the resolution. Red dashed line shows $N^2$ scaling.}
    \label{figc1}
\end{figure}
\begin{figure}[ht]
\centering
\includegraphics[width=.5\textwidth]{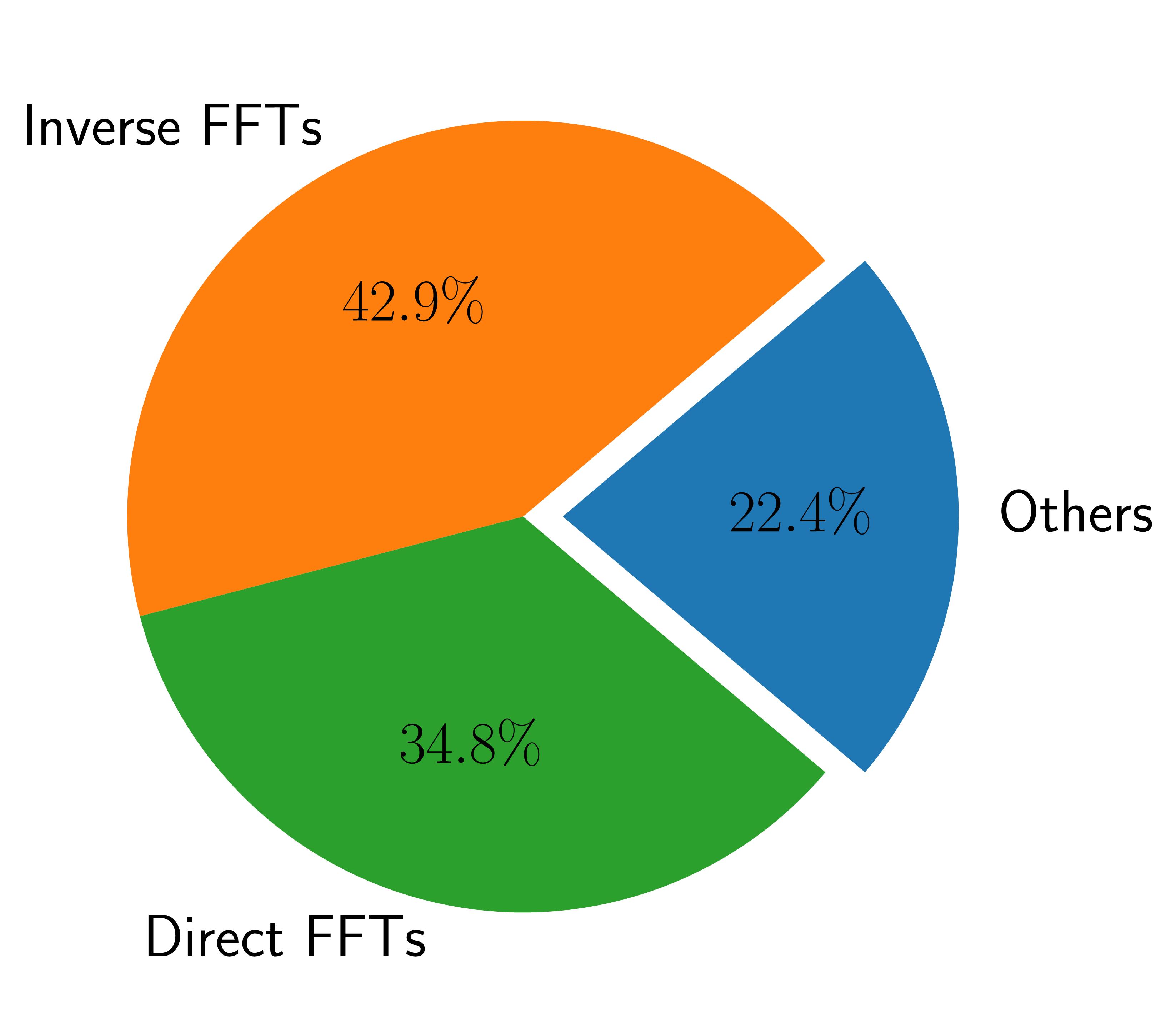}
\caption{Representative graph of the time fraction allocated to each stage in the code.}
\label{figc2}
\end{figure}

The left panel of  Figure \ref{figc1} shows the total GPU memory usage in Gb while the right panel shows computational time, both as functions of resolution $N$. For moderate resolution $N \lesssim 2000$ the memory usage is almost independent of the resolution since most of the memory is used to store the libraries, the kernel, and the resolution-independent variables. For larger resolutions, the memory used to store the 2D fields dominates and therefore it is proportional to $N^2$. We also observe a similar behaviour for the mean elapsed time. This can be explained by the relative smallness of the problem compared to the GPU parallelisation capacity. Indeed, not all the registers on the GPUs are required to fully parallelise the computation, therefore, increasing the resolution just occupies free registers not increasing the simulation time. For larger resolution, the computational time grows proportionally to the amount of computation required for the time step, i.e. to $N^2$.

Figure \ref{figc2} shows the percentage of time spent on the simulation for each RK cycle at the maximum resolution $(N^2=32768^2)$. One should note that the most computationally intensive part is due to the forward and backward FFTs that account for more than $75\%$ of the computational time. However, the importance of the forward and backward transforms is different since their subroutines are called with different frequencies. Besides, we decided to move the normalization to the forward transform since it has fewer calls per timestep. Although the integrator stability depends intrinsically on the physical properties of the system in question, we observed some practical advantages of using RK4 in some tested cases for simulations with fixed physical time $T=N_t dt$, since higher order schemes can allow one to use larger timesteps.

\section{The effect of the log correction when measuring spectral correction}\label{appb}

\begin{figure}[ht]
\centering
\includegraphics[width=.68\textwidth]{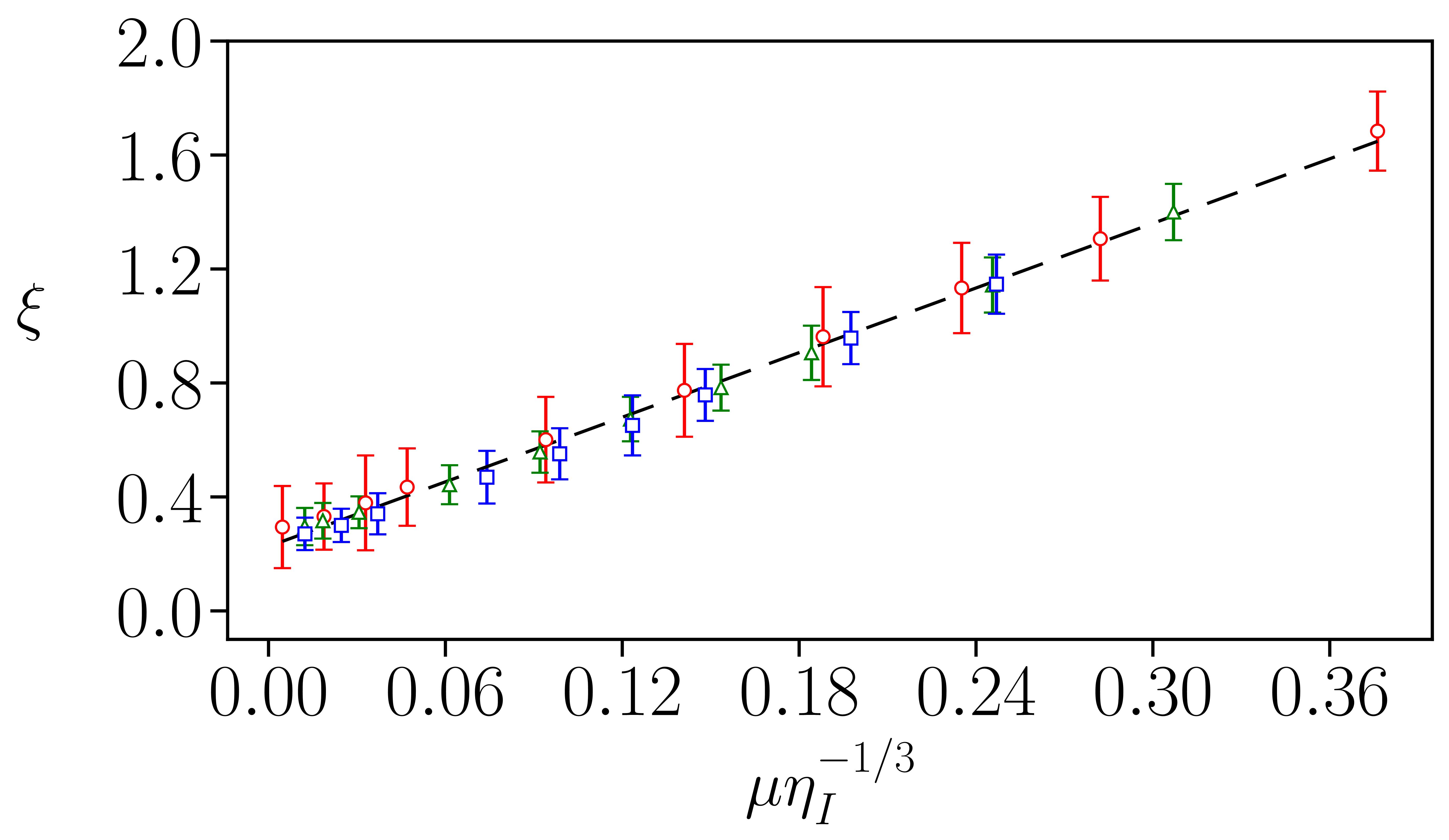}
\caption{Same as Figure~\ref{fig4} with $\xi$ fitted directly from the spectra.
The dashed line represents the relation $\xi=a\big(\mu\eta_I^{-1/3}\big)+b$
where $a=3.8\pm0.5$ and $b=0.22\pm0.08$.}
\label{figb1}
\end{figure}

To measure the correction $\xi(\mu)$, we first analyse the spectrum $E(k)$ under the assumption of a pure power-law scaling, $E(k) \propto k^{-3-\xi}$. The result is shown in Figure~\ref{figb1} where we observe a vertical shift in the y-axis which is incompatible with the arguments put forward in Sec.~\ref{sec2}. In particular, the limit $\xi(\mu\rightarrow0)\rightarrow0$ is completely missed even when the error bars are huge, which is the case of low-resolution simulations. 

\begin{figure}[ht]
\centering
\includegraphics[width=.68\textwidth]{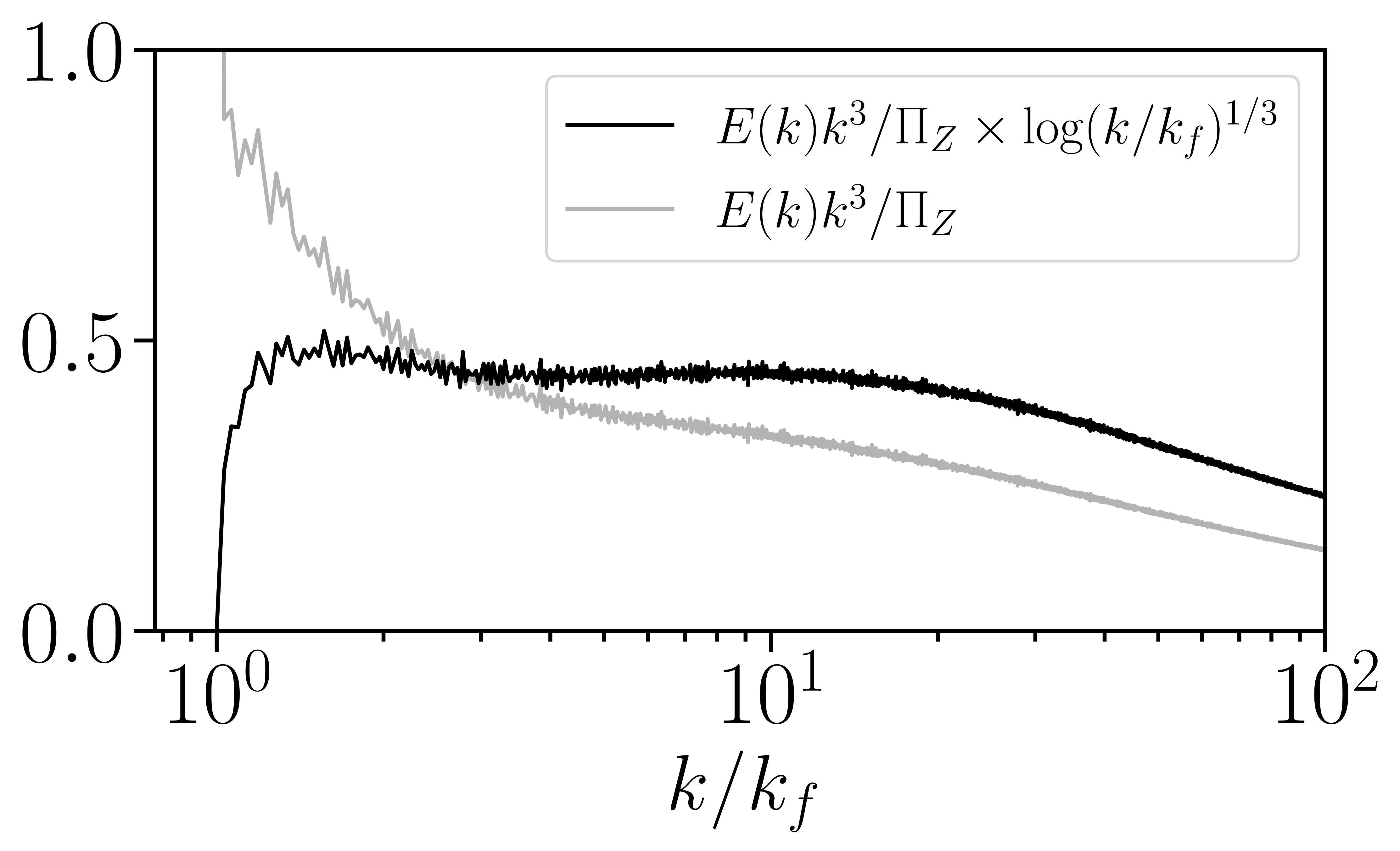}
\caption{Test for the dimensional relation \eqref{eq2.6} for a simulation on Run C with $\mu\eta_I^{-1/3}\approx0.04$.}
\label{figb2}
\end{figure}

We tested the validity of Equation~\eqref{eq2.6} for the case where $\lambda_k=\lambda_{k_f}$. This equation predicts $E(k)k^{3}/\Pi_Z(k)\approx \text{const.}$ in the enstrophy inertial range. Figure~\ref{figb2} shows this relation as functions of the wavenumber $k$ for a simulation with a small value of $\mu\eta_I^{-1/3}$. The darker curve includes the log correction term $\ln(k/k_f)^{1/3}$ as in Equation~\eqref{eq2.9} while the lighter curve shows simply Equation~\eqref{eq2.6}. By Figure~\ref{figb2}, one should note that for small friction, there exists an emergent logarithmic correction to the deformation frequency which is the source of our difficulties in fitting the correct scaling exponent through the spectrum. Indeed, if one fits the spectrum taking into account the correction the offset vanishes (not shown).
However, this procedure cannot be systematically applied for all values of $\mu$ since we expect the logarithmic correction to be less pronounced for large friction. Then, we decided to extract the correction directly from the flux $\Pi_Z(k)$ since it is not supposed to present the logarithmic term. This procedure also showed to reduce error bars for all simulations (see Figs.~\ref{fig4} and~\ref{figb1}).

\bibliography{biblio.bib}

\begin{thebibliography}{26}
\expandafter\ifx\csname natexlab\endcsname\relax\def\natexlab#1{#1}\fi
\expandafter\ifx\csname bibnamefont\endcsname\relax
  \def\bibnamefont#1{#1}\fi
\expandafter\ifx\csname bibfnamefont\endcsname\relax
  \def\bibfnamefont#1{#1}\fi
\expandafter\ifx\csname citenamefont\endcsname\relax
  \def\citenamefont#1{#1}\fi
\expandafter\ifx\csname url\endcsname\relax
  \def\url#1{\texttt{#1}}\fi
\expandafter\ifx\csname urlprefix\endcsname\relax\def\urlprefix{URL }\fi
\providecommand{\bibinfo}[2]{#2}
\providecommand{\eprint}[2][]{\url{#2}}

\bibitem[{\citenamefont{Vallis}(2017)}]{vallis2017atmospheric}
\bibinfo{author}{\bibfnamefont{G.~K.} \bibnamefont{Vallis}},
  \emph{\bibinfo{title}{Atmospheric and oceanic fluid dynamics}}
  (\bibinfo{publisher}{Cambridge University Press, Cambridge},
  \bibinfo{year}{2017}).

\bibitem[{\citenamefont{Lapeyre and Klein}(2006)}]{lapeyre2006dynamics}
\bibinfo{author}{\bibfnamefont{G.}~\bibnamefont{Lapeyre}} \bibnamefont{and}
  \bibinfo{author}{\bibfnamefont{P.}~\bibnamefont{Klein}}, \bibinfo{journal}{J.
  Phys. Oceanogr.} \textbf{\bibinfo{volume}{36}}, \bibinfo{pages}{165}
  (\bibinfo{year}{2006}).

\bibitem[{\citenamefont{Siegelman et~al.}(2022)\citenamefont{Siegelman, Klein,
  Ingersoll, Ewald, Young, Bracco, Mura, Adriani, Grassi, Plainaki
  et~al.}}]{siegelman2022moist}
\bibinfo{author}{\bibfnamefont{L.}~\bibnamefont{Siegelman}},
  \bibinfo{author}{\bibfnamefont{P.}~\bibnamefont{Klein}},
  \bibinfo{author}{\bibfnamefont{A.~P.} \bibnamefont{Ingersoll}},
  \bibinfo{author}{\bibfnamefont{S.~P.} \bibnamefont{Ewald}},
  \bibinfo{author}{\bibfnamefont{W.~R.} \bibnamefont{Young}},
  \bibinfo{author}{\bibfnamefont{A.}~\bibnamefont{Bracco}},
  \bibinfo{author}{\bibfnamefont{A.}~\bibnamefont{Mura}},
  \bibinfo{author}{\bibfnamefont{A.}~\bibnamefont{Adriani}},
  \bibinfo{author}{\bibfnamefont{D.}~\bibnamefont{Grassi}},
  \bibinfo{author}{\bibfnamefont{C.}~\bibnamefont{Plainaki}},
  \bibnamefont{et~al.}, \bibinfo{journal}{Nat. Phys.}
  \textbf{\bibinfo{volume}{18}}, \bibinfo{pages}{357} (\bibinfo{year}{2022}).

\bibitem[{\citenamefont{Juckes}(1994)}]{juckes1994quasigeostrophic}
\bibinfo{author}{\bibfnamefont{M.}~\bibnamefont{Juckes}}, \bibinfo{journal}{J.
  Atmos. Sci.} \textbf{\bibinfo{volume}{51}}, \bibinfo{pages}{2756}
  (\bibinfo{year}{1994}).

\bibitem[{\citenamefont{Xia et~al.}(2009)\citenamefont{Xia, Shats, and
  Falkovich}}]{xia2009spectrally}
\bibinfo{author}{\bibfnamefont{H.}~\bibnamefont{Xia}},
  \bibinfo{author}{\bibfnamefont{M.}~\bibnamefont{Shats}}, \bibnamefont{and}
  \bibinfo{author}{\bibfnamefont{G.}~\bibnamefont{Falkovich}},
  \bibinfo{journal}{Phys. Fluids} \textbf{\bibinfo{volume}{21}},
  \bibinfo{pages}{125101} (\bibinfo{year}{2009}).

\bibitem[{\citenamefont{Benavides and Alexakis}(2017)}]{benavides2017critical}
\bibinfo{author}{\bibfnamefont{S.~J.} \bibnamefont{Benavides}}
  \bibnamefont{and} \bibinfo{author}{\bibfnamefont{A.}~\bibnamefont{Alexakis}},
  \bibinfo{journal}{J. Fluid Mech.} \textbf{\bibinfo{volume}{822}},
  \bibinfo{pages}{364} (\bibinfo{year}{2017}).

\bibitem[{\citenamefont{Musacchio and Boffetta}(2017)}]{musacchio2017split}
\bibinfo{author}{\bibfnamefont{S.}~\bibnamefont{Musacchio}} \bibnamefont{and}
  \bibinfo{author}{\bibfnamefont{G.}~\bibnamefont{Boffetta}},
  \bibinfo{journal}{Phys. Fluids} \textbf{\bibinfo{volume}{29}},
  \bibinfo{pages}{111106} (\bibinfo{year}{2017}).

\bibitem[{\citenamefont{Musacchio and
  Boffetta}(2019)}]{musacchio2019condensate}
\bibinfo{author}{\bibfnamefont{S.}~\bibnamefont{Musacchio}} \bibnamefont{and}
  \bibinfo{author}{\bibfnamefont{G.}~\bibnamefont{Boffetta}},
  \bibinfo{journal}{Phys. Rev. Fluids} \textbf{\bibinfo{volume}{4}},
  \bibinfo{pages}{022602} (\bibinfo{year}{2019}).

\bibitem[{\citenamefont{Zhu et~al.}(2023)\citenamefont{Zhu, Xie, Xia
  et~al.}}]{zhu2023circulation}
\bibinfo{author}{\bibfnamefont{H.-Y.} \bibnamefont{Zhu}},
  \bibinfo{author}{\bibfnamefont{J.-H.} \bibnamefont{Xie}},
  \bibinfo{author}{\bibfnamefont{K.-Q.} \bibnamefont{Xia}},
  \bibnamefont{et~al.}, \bibinfo{journal}{Phys. Rev. Lett.}
  \textbf{\bibinfo{volume}{130}}, \bibinfo{pages}{214001}
  (\bibinfo{year}{2023}).

\bibitem[{\citenamefont{Moffatt}(1978)}]{moffatt1978magnetic}
\bibinfo{author}{\bibfnamefont{H.~K.} \bibnamefont{Moffatt}},
  \emph{\bibinfo{title}{Magnetic field generation in electrically conducting
  fluids}} (\bibinfo{publisher}{Cambridge University Press, Cambridge},
  \bibinfo{year}{1978}).

\bibitem[{\citenamefont{Kolmogorov}(1941)}]{kolmogorov1941local}
\bibinfo{author}{\bibfnamefont{A.~N.} \bibnamefont{Kolmogorov}},
  \bibinfo{journal}{Dokl. Akad. Nauk SSSR} \textbf{\bibinfo{volume}{30}},
  \bibinfo{pages}{301} (\bibinfo{year}{1941}).

\bibitem[{\citenamefont{Frisch}(1995)}]{frisch1995turbulence}
\bibinfo{author}{\bibfnamefont{U.}~\bibnamefont{Frisch}},
  \emph{\bibinfo{title}{\textit{Turbulence: the legacy of A.N. Kolmogorov}}}
  (\bibinfo{publisher}{Cambridge University Press, Cambridge},
  \bibinfo{year}{1995}).

\bibitem[{\citenamefont{Kraichnan}(1971)}]{kraichnan1971inertial}
\bibinfo{author}{\bibfnamefont{R.~H.} \bibnamefont{Kraichnan}},
  \bibinfo{journal}{J. Fluid Mech.} \textbf{\bibinfo{volume}{47}},
  \bibinfo{pages}{525} (\bibinfo{year}{1971}).

\bibitem[{\citenamefont{Boffetta and Ecke}(2012)}]{boffetta2012two}
\bibinfo{author}{\bibfnamefont{G.}~\bibnamefont{Boffetta}} \bibnamefont{and}
  \bibinfo{author}{\bibfnamefont{R.~E.} \bibnamefont{Ecke}},
  \bibinfo{journal}{Annu. Rev. Fluid Mech.} \textbf{\bibinfo{volume}{44}},
  \bibinfo{pages}{427} (\bibinfo{year}{2012}).

\bibitem[{\citenamefont{Rivera and Wu}(2000)}]{rivera2000external}
\bibinfo{author}{\bibfnamefont{M.}~\bibnamefont{Rivera}} \bibnamefont{and}
  \bibinfo{author}{\bibfnamefont{X.-L.} \bibnamefont{Wu}},
  \bibinfo{journal}{Phys. Rev. Lett.} \textbf{\bibinfo{volume}{85}},
  \bibinfo{pages}{976} (\bibinfo{year}{2000}).

\bibitem[{\citenamefont{Kraichnan}(1967)}]{kraichnan1967inertial}
\bibinfo{author}{\bibfnamefont{R.~H.} \bibnamefont{Kraichnan}},
  \bibinfo{journal}{Phys. Rev. Fluids} \textbf{\bibinfo{volume}{10}},
  \bibinfo{pages}{1417} (\bibinfo{year}{1967}).

\bibitem[{\citenamefont{Leith}(1968)}]{leith1968diffusion}
\bibinfo{author}{\bibfnamefont{C.~E.} \bibnamefont{Leith}},
  \bibinfo{journal}{Phys. Rev. Fluids} \textbf{\bibinfo{volume}{11}},
  \bibinfo{pages}{671} (\bibinfo{year}{1968}).

\bibitem[{\citenamefont{Batchelor}(1969)}]{batchelor1969computation}
\bibinfo{author}{\bibfnamefont{G.~K.} \bibnamefont{Batchelor}},
  \bibinfo{journal}{Phys. Fluids} \textbf{\bibinfo{volume}{12}},
  \bibinfo{pages}{II} (\bibinfo{year}{1969}).

\bibitem[{\citenamefont{Chertkov et~al.}(2007)\citenamefont{Chertkov,
  Connaughton, Kolokolov, and Lebedev}}]{chertkov2007dynamics}
\bibinfo{author}{\bibfnamefont{M.}~\bibnamefont{Chertkov}},
  \bibinfo{author}{\bibfnamefont{C.}~\bibnamefont{Connaughton}},
  \bibinfo{author}{\bibfnamefont{I.}~\bibnamefont{Kolokolov}},
  \bibnamefont{and} \bibinfo{author}{\bibfnamefont{V.}~\bibnamefont{Lebedev}},
  \bibinfo{journal}{Phys. Rev. Lett.} \textbf{\bibinfo{volume}{99}},
  \bibinfo{pages}{084501} (\bibinfo{year}{2007}).

\bibitem[{\citenamefont{Nam et~al.}(2000)\citenamefont{Nam, Ott, Antonsen~Jr,
  and Guzdar}}]{nam2000lagrangian}
\bibinfo{author}{\bibfnamefont{K.}~\bibnamefont{Nam}},
  \bibinfo{author}{\bibfnamefont{E.}~\bibnamefont{Ott}},
  \bibinfo{author}{\bibfnamefont{T.~M.} \bibnamefont{Antonsen~Jr}},
  \bibnamefont{and} \bibinfo{author}{\bibfnamefont{P.~N.}
  \bibnamefont{Guzdar}}, \bibinfo{journal}{Phys. Rev. Lett.}
  \textbf{\bibinfo{volume}{84}}, \bibinfo{pages}{5134} (\bibinfo{year}{2000}).

\bibitem[{\citenamefont{Bernard}(2000)}]{bernard2000influence}
\bibinfo{author}{\bibfnamefont{D.}~\bibnamefont{Bernard}},
  \bibinfo{journal}{Europhys. Lett.} \textbf{\bibinfo{volume}{50}},
  \bibinfo{pages}{333} (\bibinfo{year}{2000}).

\bibitem[{\citenamefont{Boffetta et~al.}(2002)\citenamefont{Boffetta, Celani,
  Musacchio, and Vergassola}}]{boffetta2002intermittency}
\bibinfo{author}{\bibfnamefont{G.}~\bibnamefont{Boffetta}},
  \bibinfo{author}{\bibfnamefont{A.}~\bibnamefont{Celani}},
  \bibinfo{author}{\bibfnamefont{S.}~\bibnamefont{Musacchio}},
  \bibnamefont{and}
  \bibinfo{author}{\bibfnamefont{M.}~\bibnamefont{Vergassola}},
  \bibinfo{journal}{Phys. Rev. E} \textbf{\bibinfo{volume}{66}},
  \bibinfo{pages}{026304} (\bibinfo{year}{2002}).

\bibitem[{\citenamefont{Boffetta et~al.}(2005)\citenamefont{Boffetta, Cenedese,
  Espa, and Musacchio}}]{boffetta2005effects}
\bibinfo{author}{\bibfnamefont{G.}~\bibnamefont{Boffetta}},
  \bibinfo{author}{\bibfnamefont{A.}~\bibnamefont{Cenedese}},
  \bibinfo{author}{\bibfnamefont{S.}~\bibnamefont{Espa}}, \bibnamefont{and}
  \bibinfo{author}{\bibfnamefont{S.}~\bibnamefont{Musacchio}},
  \bibinfo{journal}{Europhys. Lett.} \textbf{\bibinfo{volume}{71}},
  \bibinfo{pages}{590} (\bibinfo{year}{2005}).

\bibitem[{\citenamefont{Pierrehumbert et~al.}(1994)\citenamefont{Pierrehumbert,
  Held, and Swanson}}]{pierrehumbert1994spectra}
\bibinfo{author}{\bibfnamefont{R.~T.} \bibnamefont{Pierrehumbert}},
  \bibinfo{author}{\bibfnamefont{I.~M.} \bibnamefont{Held}}, \bibnamefont{and}
  \bibinfo{author}{\bibfnamefont{K.~L.} \bibnamefont{Swanson}},
  \bibinfo{journal}{Chaos Solit. Fract.} \textbf{\bibinfo{volume}{4}},
  \bibinfo{pages}{1111} (\bibinfo{year}{1994}).

\bibitem[{\citenamefont{Haugen and Brandenburg}(2004)}]{haugen2004inertial}
\bibinfo{author}{\bibfnamefont{N.~E.~L.} \bibnamefont{Haugen}}
  \bibnamefont{and}
  \bibinfo{author}{\bibfnamefont{A.}~\bibnamefont{Brandenburg}},
  \bibinfo{journal}{Phys. Rev. E} \textbf{\bibinfo{volume}{70}},
  \bibinfo{pages}{026405} (\bibinfo{year}{2004}).

\bibitem[{\citenamefont{Frisch et~al.}(2008)\citenamefont{Frisch, Kurien,
  Pandit, Pauls, Ray, Wirth, and Zhu}}]{frisch2008hyperviscosity}
\bibinfo{author}{\bibfnamefont{U.}~\bibnamefont{Frisch}},
  \bibinfo{author}{\bibfnamefont{S.}~\bibnamefont{Kurien}},
  \bibinfo{author}{\bibfnamefont{R.}~\bibnamefont{Pandit}},
  \bibinfo{author}{\bibfnamefont{W.}~\bibnamefont{Pauls}},
  \bibinfo{author}{\bibfnamefont{S.~S.} \bibnamefont{Ray}},
  \bibinfo{author}{\bibfnamefont{A.}~\bibnamefont{Wirth}}, \bibnamefont{and}
  \bibinfo{author}{\bibfnamefont{J.-Z.} \bibnamefont{Zhu}},
  \bibinfo{journal}{Phys. Rev. Lett.} \textbf{\bibinfo{volume}{101}},
  \bibinfo{pages}{144501} (\bibinfo{year}{2008}).

\end{thebibliography}

\end{document}